# Leveraging sinusoidal representation networks to predict fMRI signals from EEG

Yamin Li[a], Ange Lou[b], Ziyuan Xu[a], Shiyu Wang[c], Catie Chang[a,b,c]

[a]Department of Computer Science, Vanderbilt University, Nashville, TN, USA
[b]Department of Electrical and Computer Engineering, Vanderbilt University, Nashville, TN, USA
[c]Department of Biomedical Engineering, Vanderbilt University, Nashville, TN, USA
{yamin.li, ange.lou, ziyuan.xu, shiyu.wang.1, catie.chang}@vanderbilt.edu

**ABSTRACT**

In modern neuroscience, functional magnetic resonance imaging (fMRI) has been a crucial and irreplaceable tool that provides a non-invasive window into the dynamics of whole-brain activity. Nevertheless, fMRI is limited by hemodynamic blurring as well as high cost, immobility, and incompatibility with metal implants. Electroencephalography (EEG) is complementary to fMRI and can directly record the cortical electrical activity at high temporal resolution, but has more limited spatial resolution and is unable to recover information about deep subcortical brain structures. The ability to obtain fMRI information from EEG would enable cost-effective, naturalistic imaging across a wider set of brain regions. Further, beyond augmenting the capabilities of EEG, cross-modality models would facilitate the interpretation of fMRI signals. However, as both EEG and fMRI are high-dimensional and prone to noise and artifacts, it is currently challenging to model fMRI from EEG. Indeed, although correlations between these two modalities have been widely investigated, few studies have successfully used EEG to directly reconstruct fMRI time series. To address this challenge, we propose a novel architecture that can predict fMRI signals directly from multi-channel EEG without explicit feature engineering. Our model achieves this by implementing a Sinusoidal Representation Network (SIREN) to learn frequency information in brain dynamics from EEG, which serves as the input to a subsequent encoder-decoder to effectively reconstruct the fMRI signal from a specific brain region. We evaluate our model using a simultaneous EEG-fMRI dataset with 8 subjects and investigate its potential for predicting subcortical fMRI signals. The present results reveal that our model outperforms a recent state-of-the-art model, and indicates the potential of leveraging periodic activation functions in deep neural networks to model functional neuroimaging data.

## 1. DESCRIPTION OF PURPOSE

As the two most frequently used non-invasive neuroimaging modalities, functional magnetic resonance imaging (fMRI) and electroencephalography (EEG) play essential roles in advancing our understanding of the human brain and its complexities. The simultaneous recording and analysis of these two modalities have also gained substantial attention, owing to their complementary strengths and physiological views of brain functioning[1]. Though fMRI helps to more precisely locate the source of neural activities, it suffers from hemodynamic blurring that introduces uncertainty in the timing of neural activity, which can be more precisely achieved by EEG. Moreover, the high cost and incompatibility with metal implants also hinder MRI from some application scenarios. A number of existing studies have tried to build bridges between the two modalities, such as through correlational or machine learning frameworks[2,3]. However, factors such as the high dimensionality, complexity, and low signal-to-noise ratio of both EEG and fMRI data create challenges for accurately modeling the translation between EEG and fMRI and for directly reconstructing fMRI time series from EEG.

The rapid development of deep learning in the past decades has facilitated multimodal learning and cross-modal prediction. Recently, Kovalev et al.[4] proposed a deep learning framework to predict subcortical fMRI signals from 30-channel EEG signals. While this work contributed a major step toward the feasibility of EEG-fMRI translation, the median performances were correlations on the order of 0.3-0.5, leaving room for performance improvement. One potential limitation of this model is that it is not explicitly designed to represent frequency features from EEG; yet, prior work has noted that resolving spectral information in the EEG may yield better prediction accuracy than that obtained without considering the frequency-band distribution of EEG[3,4]. In recent years, the use of periodic activation functions, specifically the sine function, has been gaining traction because of its excellent performance in implicit neural representations[5]. By mapping the periodic patterns and continuous functions underlying a signal, this sinusoidal representation network (SIREN) is proving to be well-suited for representing complex natural signals and their derivatives[5,6]. Therefore, we hypothesize that SIREN would also be effective in learning frequency information in EEG time series, and could thereby significantly improve the prediction of fMRI from EEG. Here, we propose a novel framework that embeds SIREN into a deep learning model with an EEG feature encoder and fMRI decoder, aiming to reconstruct fMRI signals directly from EEG raw data without explicit feature engineering. This method provides a promising avenue for learning interpretable (non-abstract) representations from these two neuroimaging modalities, and improving understanding of the relationship between them.

## 2. METHODS

### 2.1 Dataset and preprocessing

#### 2.1.1 Dataset

Simultaneous resting-state "Eyes Open – Eyes Closed" EEG-fMRI data from 8 subjects[7] were used in this analysis. The fMRI acquisition parameters were: TR=1.95s for the first 4 subjects, TR=2.00s for other four subjects; resolution=3mm isotropic; duration=5min per subject. The EEG was recorded using a 30-channel MR-compatible electrode cap with a sampling frequency of 5000 Hz and was then corrected for gradient artifacts and down-sampled to 1000 Hz. Other detailed information about the dataset can be found in the related paper[7].

#### 2.1.2 fMRI and EEG preprocessing

The clean EEG data were filtered into 1-100Hz, and notch filtered at 50, 100, and 150 Hz to remove the power-line signal. The EEG signals were then re-referenced to the average reference. The fMRI signals were slice-timing corrected and spatially registered to standard MNI space, and a spatial Gaussian filter with Full-Width at Half Maximum (FWHM) = 3 mm was applied to increase the signal-to-noise ratio (SNR). Then, confound regression was carried out to remove motion-related artifacts and slow trends due to scanner drifts. Subsequently, the fMRI signals in several regions of interest (ROIs) were extracted using the Harvard-Oxford structural atlas[8]. In this study, we focused on four bilaterally symmetric basal ganglia regions: pallidum, caudate, putamen, and accumbens. The preprocessed subcortical fMRI signals were interpolated to 100 Hz, and the EEG was downsampled to the same sampling rate. Since the hemodynamic response measured by fMRI is slower and delayed compared to the actual onset of neural activity, we shifted the EEG signals with a time delay that approximates that of the hemodynamic response function (HRF). We set the time delay as 6 seconds, as this value attained the best performance in the previous experiments conducted by Kovalev et al.[4]. In our analysis, we trained subject-specific models given the potentially unique response properties of individuals. The preprocessed data for each subject were divided into training and testing sets in a ratio of 4:1, i.e., 4 minutes for training and 1 minute for testing. The data are further divided into windows of length $t_{win}$=20.48 seconds (with the time shift already incorporated into the EEG data to accommodate the HRF delay). To form the training sample pair $\{X_j, y_j\}$ for the $j^{th}$ window, we extract a window of EEG data formed at a randomly selected staring time index $t_j$ as $X_j = X(:, t_j: t_j + t_{win} * Fs)$ and a corresponding window of fMRI data as $y_j = y(:, t_j: t_j + t_{win} * Fs)$, where Fs is the sampling rate of 100 Hz.

### 2.2 Model Architecture

Our model comprises two main components: 1) sinusoidal representation network (SIREN) blocks and 2) feature encoder and decoder blocks. We first used SIREN for initial EEG frequency-related feature extraction. The output of SIREN serves as the input of the following encoder-decoder to predict the fMRI signals.

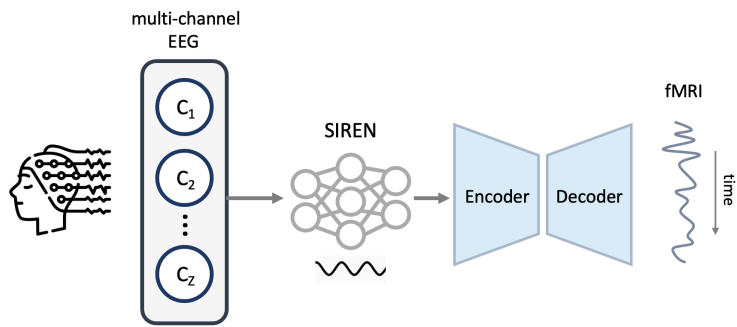

Figure 1. Model architecture

### 2.2.1 Sinusoidal Representation Network (SIREN)

Inspired by the works of Sitzmann et al.[5] and Kazemi et al.[9], we propose a framework that leverages the periodic activation function in each layer of a multilayer perceptron (MLP), i.e., the Sine layer, to extract EEG features, thereby learning common representations between EEG and fMRI without explicit feature engineering:

$$\Phi(\mathbf{x}) = \mathbf{W}_n(\phi_{n-1} \circ \phi_{n-2} \circ \dots \circ \phi_0)(\mathbf{x}) + \mathbf{b}_n, \quad \mathbf{x}_i \mapsto \phi_i(\mathbf{x}_i) = \sin(\mathbf{W}_i \mathbf{x}_i + \mathbf{b}_i), \tag{1}$$

where $\phi_i : \mathbb{R}^{M_i} \mapsto \mathbb{R}^{N_i}$ is the $i^{th}$ layer of the network. It comprises the affine transform defined by the weight matrix $\mathbf{W}_i \in \mathbb{R}^{N_i \times M_i}$ and the bias $\mathbf{b}_i \in \mathbb{R}^{N_i}$ applied on the input $\mathbf{x}_i \in \mathbb{R}^{M_i}$, followed by the sine non-linearity applied to each component of the resulting vector. This section consists of an input layer and K hidden layers (K = 1 in this study), followed by a linear projection layer.

### 2.2.2 Feature Encoder and Decoder

The SIREN output is then sent into the encoder and decoder blocks to predict the fMRI signals from each ROI. The encoder consists of N encoder blocks following the architecture of the wav2vec 2.0 model[10], followed by the dropout block[11] (N = 4, dropout rate = 0.3 for all layers in this study). Each encoder block has a down-sampling operation, i.e., max pooling in this analysis, which efficiently increases the receptive field while retaining important information. The decoder comprises the same symmetric building blocks and up-samples the latent space features to produce the fMRI signal of a certain ROI.

### 2.2.3 Learning Objective

In this training paradigm, the model optimizes the linear combination of two losses, the mean squared error (MSE) loss and the correlation loss:

$$Loss = L_{mse} + \alpha L_{corr}. \tag{2}$$

Here, $\alpha$ is a hyperparameter we tune, and for the correlation loss, we use the negative Pearson correlation coefficient:

$$\mathcal{L}_{corr} = -r(y, \hat{y}) = -\frac{\sum_{i=1}^{n}(y_i - \overline{y_i})(\hat{y}_i - \overline{\hat{y}_i})}{\sqrt{\sum_{i=1}^{n}(y_i - \overline{y_i})^2 (\hat{y}_i - \overline{\hat{y}_i})^2}}. \tag{3}$$

## 2.3 Experiment and implementation details

The proposed model was implemented using PyTorch. We chose AdamW optimizer with β1=0.9 and β2=0.999, and the correlation loss coefficient $\alpha$=0.1. The batch size is 32 and initial learning rate is 3e-4 with weight decay of 3e-4. The whole analysis was running on a single RTX A5000 GPU. In this work, to further evaluate the performance of our architecture and further explore if SIREN significantly improves the prediction of fMRI signals, we compare our model with the state-of-art EEG-fMRI prediction model BEIRA[4] and ridge regression.

## 3. RESULTS

Table 1 shows the quantitative performance of our model in all subjects and in one selected subject (the best case). The results show that our model outperforms the current state-of-art deep learning model BERIA. Figure 2 depicts the predicted and ground-truth fMRI signals for the single (best-case) subject, and indicates that our model is able to correctly capture most of the phase information of subcortical fMRI signals.

Table 1. Comparison of proposed model with existing models for subcortical fMRI time series prediction from EEG

| **All subjects** | Pallidum | Caudate | Putamen | Accumbens | Average |
|---|---|---|---|---|---|
| Ridge regression model[4] | 0.05 ± 0.15 | 0.34 ± 0.14 | 0.38 ± 0.24 | 0.26 ± 0.10 | 0.24 ± 0.11 |
| BEIRA[4] | 0.37 ± 0.04 | 0.47 ± 0.14 | 0.48 ± 0.16 | 0.42 ± 0.05 | 0.44 ± 0.05 |
| Ours | **0.43 ± 0.15** | **0.49 ± 0.12** | **0.51 ± 0.14** | **0.43 ± 0.13** | **0.47 ± 0.04** |
| **One subject** | Pallidum | Caudate | Putamen | Accumbens | Average |
| Ridge regression model[4] | 0.01 | 0.40 | 0.48 | 0.10 | 0.25 ± 0.20 |
| BEIRA[4] | 0.39 | 0.62 | 0.55 | 0.37 | 0.48 ± 0.02 |
| Ours | **0.635** | **0.70** | **0.62** | **0.50** | **0.61 ± 0.09** |

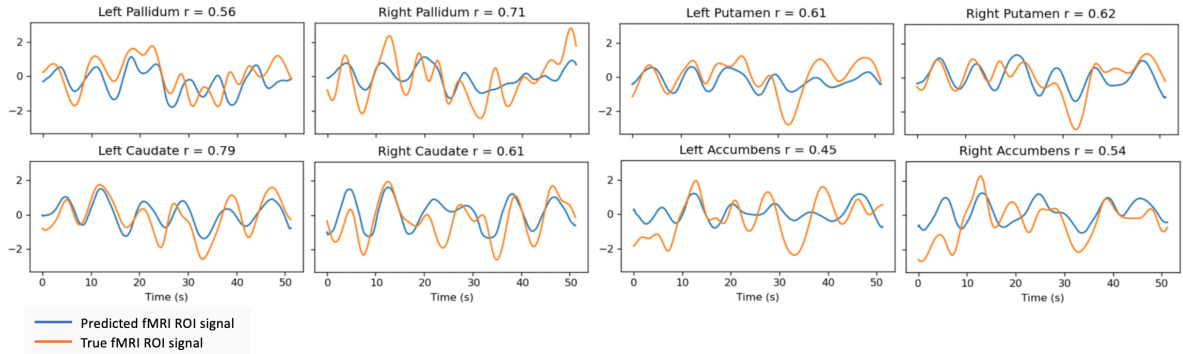

Figure 2. Real and predicted fMRI time series of different subcortical ROIs from one subject. Orange line: real signal, blue line: predicted signal.

4. **NEW OR BREAKTHROUGH WORK TO BE PRESENTED** (THIS WORK HAS NOT BEEN PRESENTED ELSEWHERE)

We propose a novel deep-learning framework to predict subcortical fMRI time series directly from EEG without explicit feature engineering. We achieve this by leveraging the periodic activation function, i.e., sine function, in MLP to form a sinusoidal representation network (SIREN) that can effectively extract complex frequency information from EEG. The features extracted by SIREN effectively facilitate the prediction performance of CNN and achieved the highest accuracy.

5. **CONCLUSION**

Our proposed model successfully reconstructs several subcortical fMRI signals from EEG time series and significantly improves the prediction accuracy compared with existing models. This work contributes a novel framework that uses periodic activation function in deep neural networks to learn representations of functional neuroimaging data. Future work would try to predict more brain areas, and assess performance on different task conditions and patient populations.